Соколов Володимир Юрійович
старший викладач кафедри інформаційної та кібернетичної безпеки
Київський університет імені Бориса Грінченка, Київ, Україна
OrcID: 0000-0002-9349-7946
*v.sokolov@kubg.edu.ua*

Вовкотруб Богдан Володимирович
магістрант
Київський університет імені Бориса Грінченка, Київ, Україна
OrcID: 0000-0002-6196-4554
*bvovkotrub@gmail.com*

Зоткін Євген Олександрович
магістрант
Київський університет імені Бориса Грінченка, Київ, Україна
OrcID: 0000-0002-6029-1044
*evgeniy.zotkin199702@gmail.com*

# ПОРІВНЯЛЬНИЙ АНАЛІЗ ПРОПУСКНОЇ ЗДАТНОСТІ МАЛОПОТУЖНИХ БЕЗПРОВОДОВИХ IOT-КОМУТАТОРІВ

**Анотація.** У статті приведено дослідження та порівняльний аналіз пропускної здатності малопотужних безпроводових IoT-пристроїв у ролі безпроводових комутаторів. На прикладі досліджувалися такі IoT-пристрої: Raspberry Pi 3 Model B та Raspberry Pi Zero W. За допомогою датчиків DS18B20 та INA219 було досліджено та проаналізовано залежність швидкості передачі мультимедійних даних по FTP в безпровідній Wi-Fi мережі від впливу температури процесора комутатора, температури зовнішнього середовища та споживаних комутатором струму і напруги. Виявлено переваги датчиків з GPIO інтерфейсом над аналоговими вимірювачами для даного досліду. Значну частину роботи присвячено на розробку автоматизації отримання результатів з GPIO-інтерфейсів, що дало змогу усунути помилки людського фактору та отримати більш точні показники вимірювань. Автоматизація вимірювань була розроблена за допомогою мови програмування Python 3.7. За допомогою бібліотеки *ina219* вдалося отримати показники струму та напруги з плати INA219. Для того, щоб отримати показники температури достатньо вбудованої у Python бібліотеки для зчитування файлів з температурою у операційній системі Raspbian. У роботі приділено увагу щодо синхронності записів результатів вимірювань для більш точного аналізу. Тому був розроблений FTP-клієнт, який заміряє швидкість завантаження файлу з FTP-сервера і записує результати одночасно з вимірюванням температури, струму та напруги. Для цього приділено увагу до багатопоточності у мові програмування Python та передачі команд за допомогою TCP-сокетів даною мовою. Як результатом було розраховано залежність вимірюваних факторів за допомогою формули кореляції Пірсона. Дані фактори вимірювання впливають на автономність і споживання енергії, що для IoT-пристроїв дуже важливо, тому серед досліджуваних пристроїв були наведені рекомендації щодо їх вибору при використанні в залежності від умов.

**Ключові слова:** пропускна здатність; безпроводова мережа; IoT; INA219; DS18B20; Raspberry Pi; RPi.

## 1. ВСТУП

На сьогоднішній день широко розповсюджені IoT-пристрої, які здатні автоматизувати різні процеси без участі людини. Для таких систем як правило використовуються мікрокомп'ютери з низьким споживанням енергії і разом з ними технології безпроводової передачі інформації по мережі. Надійність і швидкість доставки інформації є важливою проблемою при її передачі в таких системах при найменшому споживанні енергії.

Одними з найбільш широко поширених мікрокомп'ютерів є Raspberry Pi (RPi). Вони використовуються в якості розумного будинку, підключення різних датчиків, в дата-центрах в якості вимірювання показань температури в серверних приміщеннях, відеоспостереження тощо. Тому, в якості дослідження використовуються RPi 3 Model B [1] та RPi Zero W [2].

**Аналіз попередніх праць.** Споживання струму і напруги є важливим питанням, адже від цього залежить довготривалість роботи при умові, якщо живлення пристрою відбувається за допомогою акумулятора, наприклад як у роботі [3], у якій проводився обмін даними між користувачем та точкою доступу у безпровідній мережі. Якщо використовувати такий приклад для обміну даних великого обсягу, то постає питання щодо їх швидкої доставки. Також збільшення обсягу даних призведе до збільшення навантаження на комутатор, а це означає що споживання енергії буде збільшеним. Тому потрібно підібрати пристрій, який буде менше споживати електроенергію і при цьому матиме максимальну пропускну здатність.

Іншою проблемою передачі даних у безпровідній мережі є відстань між клієнтом та точкою доступу. Адже при збільшенні відстані збільшується кількість завад, що призведе до зменшення швидкості і надійності передачі даних. У роботі [4] було приведено приклад що краще використовувати пристрої з екрануванням для отримання інформації. Теплопровідність у цій роботі зіграла значну роль, адже її недостатність призвела до того, що згорів один із мікроконтролерів. Тому малопотужний пристрій повинен бути більш витривалим до збільшення температури, яка впливає на пропускну здатність, що й досліджуватиметься у даній статті.

**Головним завданням** даної роботи є дослідження впливу температури, завантаження процесора IoT-комутатора, а також його споживання струму і напруги, при передачі файлів по безпроводовій Wi-Fi мережі через FTP.

## 2. ПРОГРАМНО-АПАРАТНЕ ЗАБЕЗПЕЧЕННЯ ЕКСПЕРИМЕНТУ

Для чистоти експерименту обидві машини налаштовані ідентично з однієї і тієї ж операційною системою, і тими ж налаштуваннями. Тому все, що написано, як RPi — це відноситься до RPi Zero і RPi 3. Для того, щоб була мінімальне навантаження (ОС) на систему, в ролі операційної системи використовувалася ОС без графічного інтерфейсу — *Raspbian OS Lite*. Особливістю даної ОС є відсутність «зайвих» програм, які уповільнюють роботу комп'ютера. У даній версії ОС присутній пакет для роботи з GPIO. RPi виступають в ролі безпроводової Wi-Fi точки доступу, налаштована по прикладу з [5]. На точках доступу встановлено FTP-сервер *vsftpd*, який простий в налаштуванні, швидкий і безпечний. Розгорнутий на окремій машині FTP-клієнт завантажував з RPi один і той же файл на різних дистанціях: 0,5 м і 5 м по 10 вимірювань на кожному відстані. В результаті буде враховуватися середнє значення всіх випробувань. Кожні 5 с знімаються показники: завантаження і температура серверного процесора, споживані струм і напруга сервером, а також швидкість завантаження файлу на FTP-клієнті. Принцип роботи зображений на рис. 1.

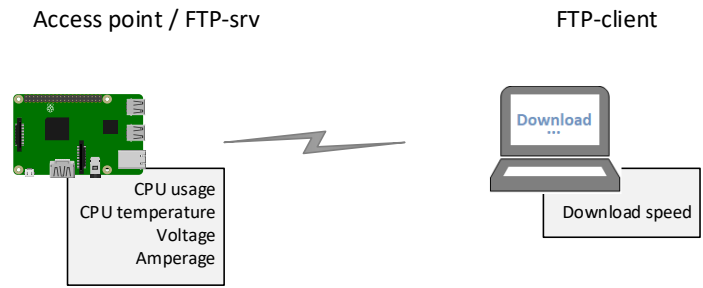

*Рис. 1. Схема експерименту*

На другому етапі було автоматизація процесу запису файлів історії. Це дозволить мінімізувати помилки людського фактору під час запису результатів, а значить підвищить точність вимірювань і швидкість запису. Для вимірювання напруги і струму використовувався цифровий датчик INA219. Крім вбудованого датчика температури процесора додатково використовується зовнішній датчик DS18B20. Обидва датчика підключаються до GPIO інтерфейсу, що дозволяє отримувати інформацію в текстовому форматі, яке використовується для автоматизації вимірювань.

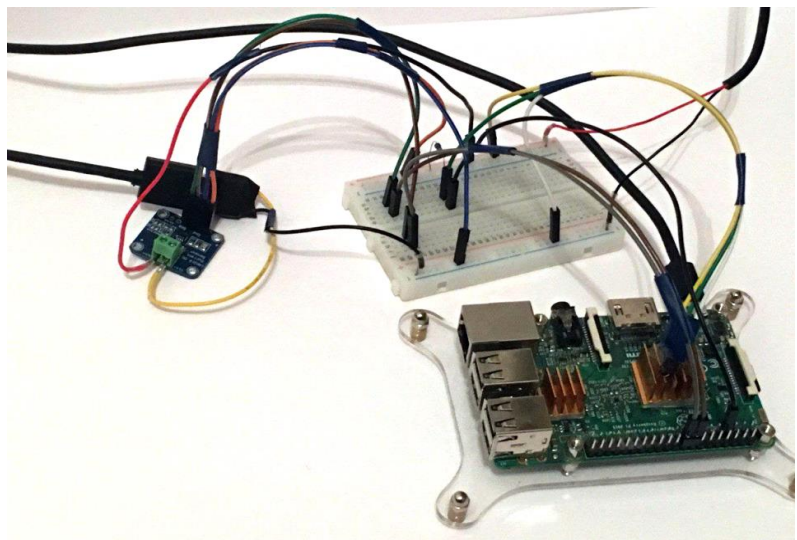

*Рис. 2. Тестовий стенд RPi 3*

## 3. ОСОБЛИВОСТІ РЕАЛІЗАЦІЇ

Спочатку планувалося проводити вимірювання напруги і струму за допомогою USB-тестера, зображеного на рис. 3.

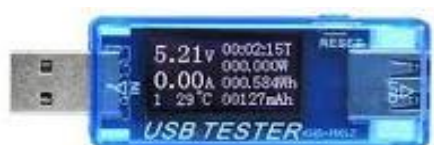

*Рис. 3. Аналоговий вимірювач напруги та струму*

Проблема USB-тестерів полягає в тому, що з них не можна знімати показники в електронному текстовому форматі. Було два варіанти вирішення даної проблеми: перший — кожні 5 с фотографувати стан датчиків тестера, а потім обробляти ці знімки, і другий варіант — використовувати датчики з GPIO інтерфейсом. Перший варіант більш ресурсозатратний, тому що на обробку знімків йде більше часу і процесорної потужності, ніж при отриманні інформації через GPIO інтерфейси RPi. Тому для експерименту обраний другий варіант.

Для того, щоб результати були точними на обох машинах, вони повинні запускатися одночасно і працювати синхронно, і зніматися в однакові проміжки часу.

Проблема була вирішена за допомогою програмного інтерфейсу забезпечення синхронного обміну даними між процесами (TCP-сокети). Процеси при такому обміні можуть виконуватися на різних ЕОМ, пов'язаних між собою мережею.

Далі буде розглянуто процес розробки автоматизації за допомогою скриптів sensors.py на серверній частини, який знімає показники з датчиків з проміжками часу по 5 с, і ftp_clt.py на клієнтській частині, який завантажує файл з FTP-сервера і записує в журнал швидкість закачування з тими ж проміжками по часу. Ці скрипти на машинах взаємодіють між собою за допомогою TCP-сокети за таким алгоритмом:
1. Сервер прослуховує порт 5555.
2. Клієнт посилає команду для «старту» на цей порт.
3. Сервер отримує команду і обидва одночасно починають вимірювання.

## 4. ОТРИМАННЯ ДАНИХ З ДАТЧИКІВ

На Server-PC запущений скрипт *sensors.py*, який отримує дані з датчиків DS18B20 і INA219. Для роботи з INA219 використовується бібліотека *ina219*.

```
from ina219 import INA219

def get_electrick():
   ina = INA219(0.1, 0.2, busnum=3)
   ina.configure(ina.RANGE_16V, ina.GAIN_AUTO)

   voltage = str('%.3f' % ina.voltage())
   ampere = str('%.3f' % (ina.current() / 1000))
   return (voltage, ampere)
```

Щоб отримати статус завантаженості процесора використовується бібліотека *psutil*.

```
import psutil

def getCPUuse():
   CPUusage = psutil.cpu_percent()
   return str(CPUusage)
```

При підключенні датчика температури DS18B20 автоматично створюється файл **/sys/bus/w1/devices/<serial_number_of_DS18B20>/w1_slave**, у якому відображається температура у текстовому форматі. Тому для того, щоб у *sensors.py* отримати температуру зовнішнього середовища, достатньо відкрити і розпарсити цей файл за допомогою вбудованої бібліотеки *os* у Python.

```
import os

def out_temp():
  try:
    out_t = os.popen("tail -1 /sys/bus/w1/devices/28-83185049eaff/
w1_slave | awk '{print $NF}'").readline().split('t=')[1].split('\n')
    out_t = float(out_t[0]) / 1000
    return str('%.1f' % out_t)
  except:
    out_t = '00.0'
    return out_t
```

Також, при ініціалізації команди **vcgencmd measure_temp**, також за допомогою *os* виводиться температура процесора з вбудованого датчика.

```
def cpu_temp():
  tempC = os.popen('vcgencmd measure_temp').readline()
  return tempC.replace('temp=','').replace("'C\n",'')
```

Функція <get_sensors()> використовує вище наведені функції, і через кожних 5 с виводить на екран інформацію з датчиків і записує її в журнал. Вона ініціалізується після того, як за допомогою бібліотеки *socket* отримає повідомлення від клієнта: «FTP is running». Блок-схема роботи <get_sensors()> відображена на рис. 4.

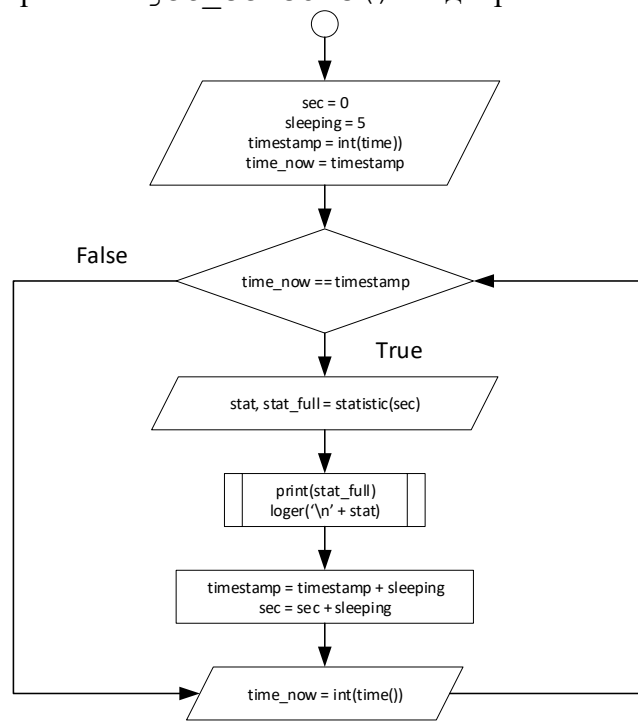

*Рис. 4. Блок-схема функції <get_sensors()>*

```
def create_socket(sock_type, HOST, PORT):
  sock = socket.socket(socket.AF_INET, socket.SOCK_STREAM)
```

```
    if sock_type == 'srv':
        sock.bind((HOST, PORT))
        sock.listen(1)
    elif sock_type == 'clt':
        sock.connect((HOST, PORT))
    else:
        sock = False
    return sock

while True:
    sock = create_socket('srv', '', 5555)
    conn, addr = sock.accept()
    message_from_ftp = conn.recv(1024)
    message_from_ftp = message_from_ftp.decode('utf-8')
    print(message_from_ftp)

    if message_from_ftp == 'download from FTP is running...':
        conn.close()
        get_sensors()
```

Загальна схема роботи скрипта *sensors.py* зображена на рис. 5.

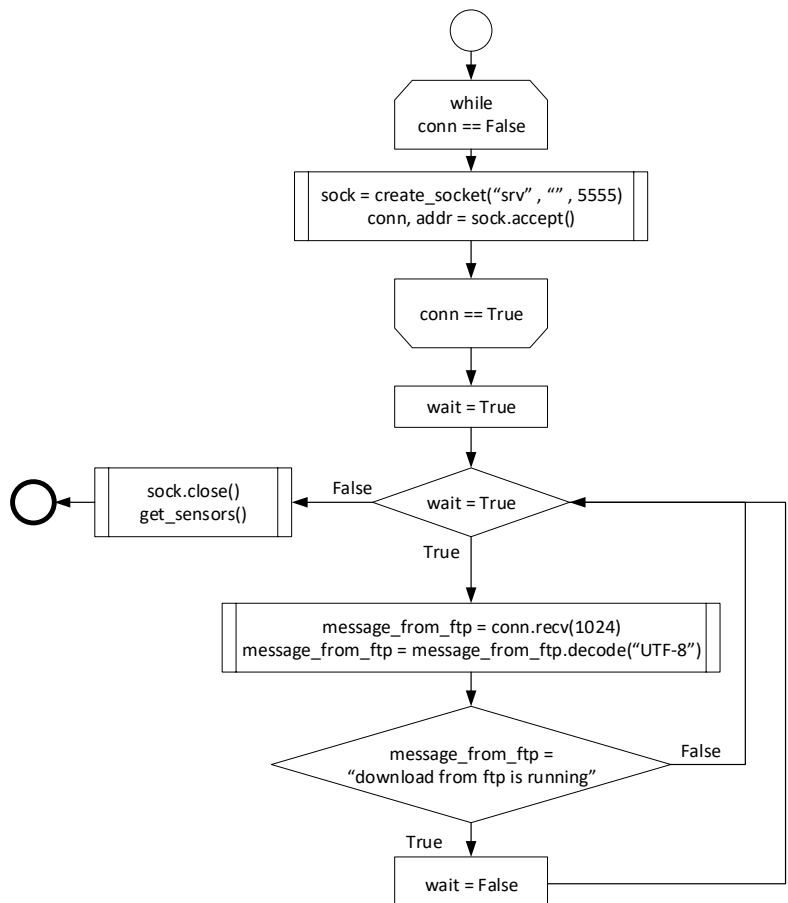

*Рис. 5. Загальна блок-схема sensors.py*

## 5. ПРОЦЕС ОТРИМАННЯ ФАЙЛУ ІЗ FTP-СЕРВЕРА

Під час запуску *ftp_clt.py* спочатку відбувається підключення до FTP-сервера, далі виконується пошук файлу, який потрібно завантажити. Для того, щоб можна було стежити за процесом завантаження файлу, потрібно отримати повний розмір файлу на сервері за допомогою `<ftp.size(file)>` — виконується порівняння розміру файлу на клієнті і на сервері.

Після підключення до FTP-сервера, *ftp_clt.py* відправляє серверу повідомлення: «FTP is runnig» за допомогою сокета на сервер. Скрипт *sensors.py* отримує дане повідомлення і разом з *ftp_clt.py* вимикають TCP-сокети і починають одночасно вимірювати показання з перериваннями 5 с.

Для того, щоб *ftp_clt.py* міг одночасно завантажувати файл за допомогою функції ftp_download(file) і паралельно вимірювати швидкість завантаження файлу за допомогою `<speedtester(file_size_ftp, file, sleeping)>`, використовується модуль *Process* з бібліотеки *multiprocessing*. Всі результати вимірювань відображаються на екрані та записуються у журнал.

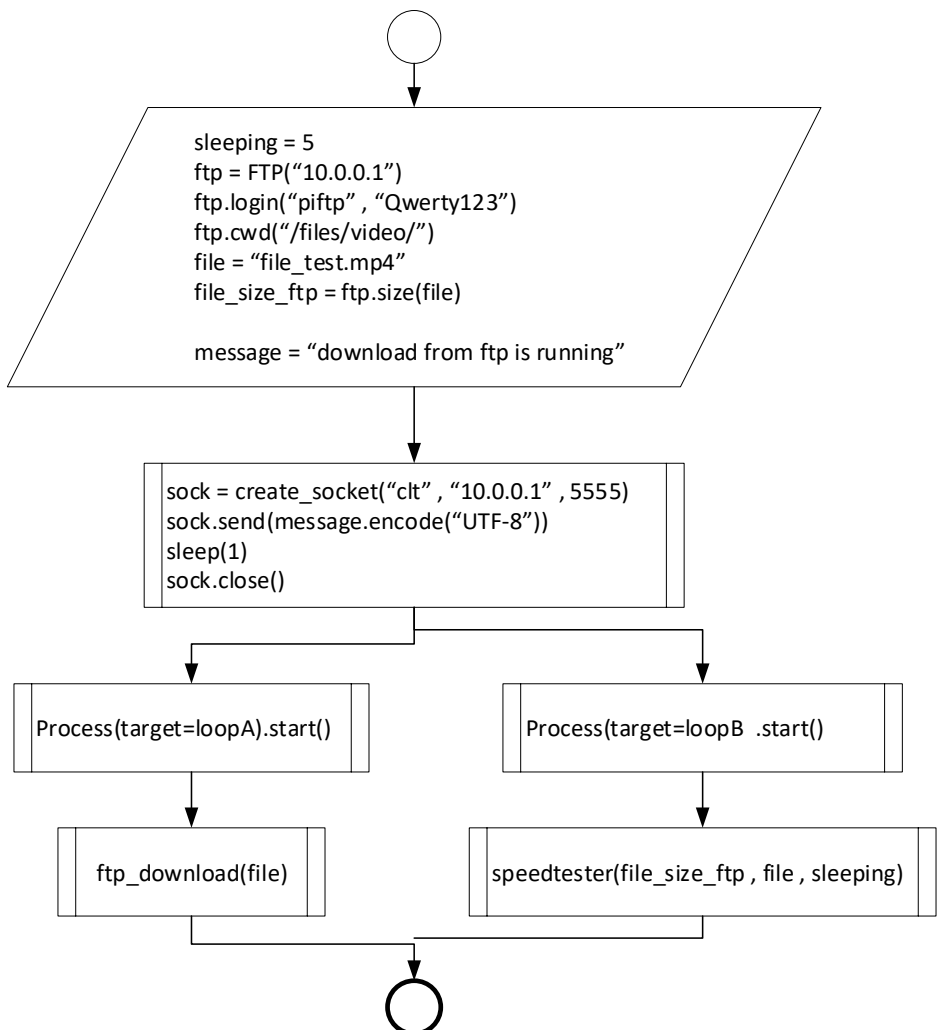

*Рис. 6. Загальна блок-схема ftp_clt.py*

```
from multiprocessing import Process

def loopA():
    ftp_download(file)

def loopB():
    speedtester(file_size_ftp, file, sleeping)

if __name__ == '__main__':
    Process(target=loopA).start()
    Process(target=loopB).start()
```

**6. РЕЗУЛЬТАТИ ВИМІРЮВАНЬ**

На відстані 0,5 м клієнта від точки доступу, завантаження файлу розміром 655 МБ відбувалася приблизно за один і той же час на обох пристроях, але на RPi 3 графік вийшов більш прямолінійним (див. рис. 7).

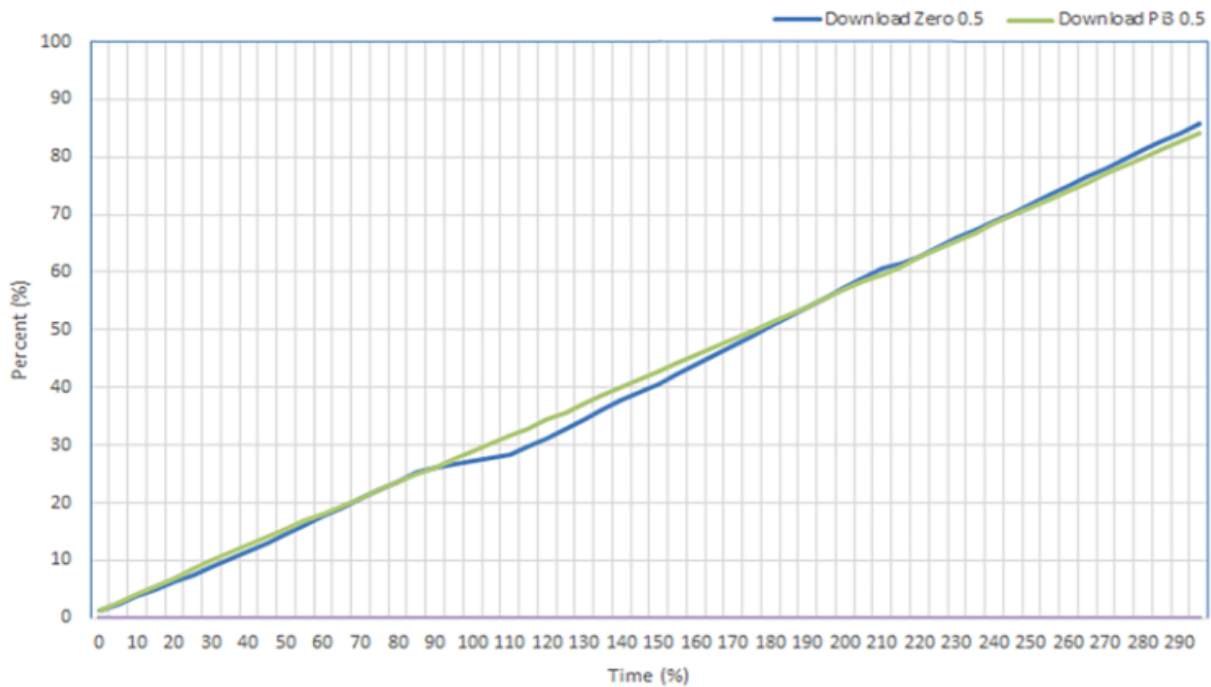

*Рис. 7. Порівняння статусу завантаження файлу за часом на відстані 0,5 м*

На рис. 8 помітно, що на довшій відстані завантаження файлу з RPi Zero відбулося швидше ніж RPi 3.

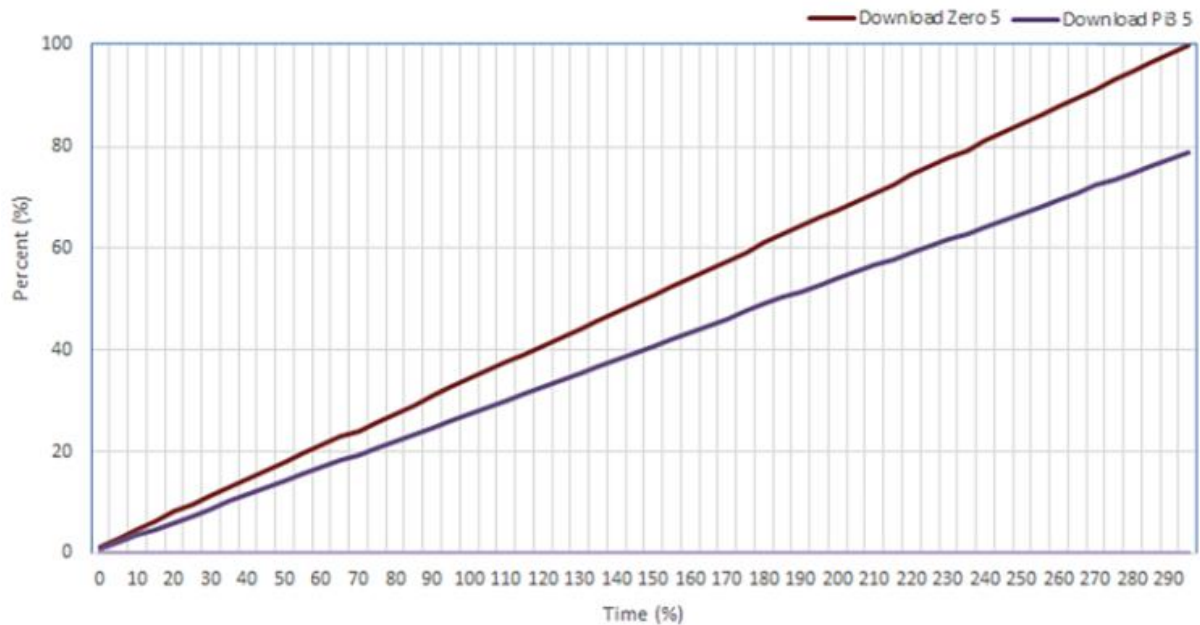

*Рис. 8. Порівняння статусу завантаження файлу за часом на відстані 5 м*

Процесор на RPi 3 був менш завантажений ніж на RPi Zero, тому на RPi 3 можна запускати більше ресурсів ніж на RPi Zero. Цікавим фактом є те, що на більш довгій відстані клієнта від точки доступу, завантаження процесора RPi 3 була меншим, а RPi Zero на більш довгому відстані періодично сильно зменшувалася, але прагнула до максимального значення (рис. 9).

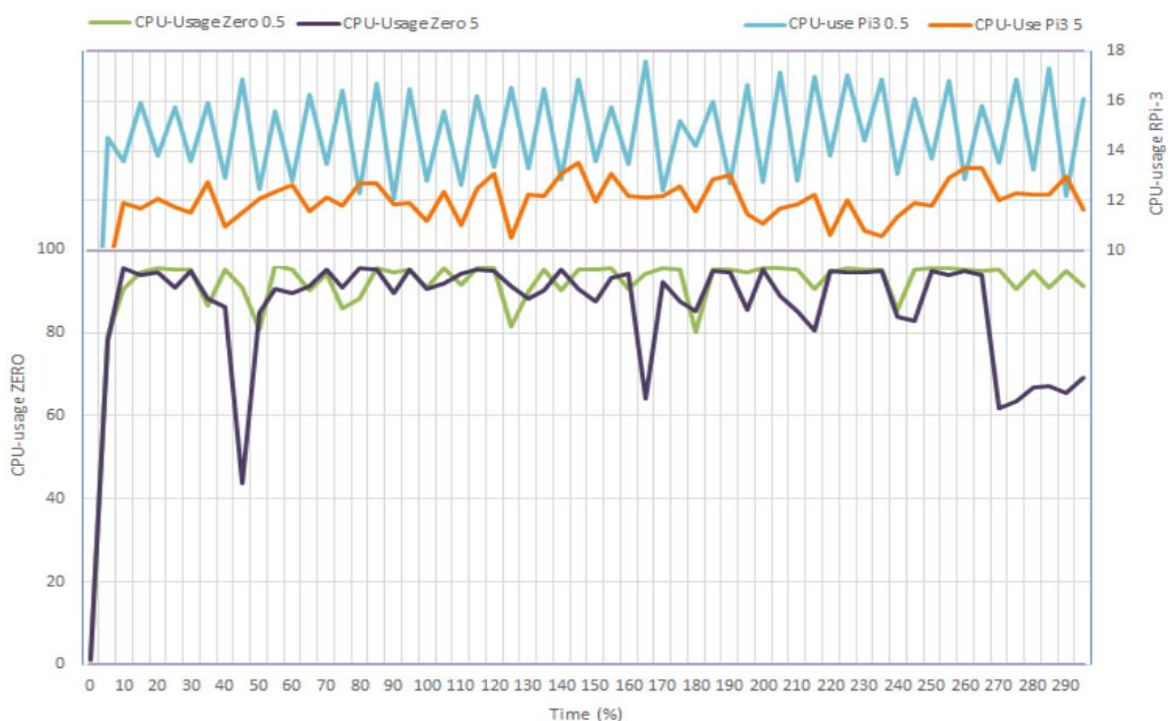

*Рис. 9. Порівняння завантаження процесора RPi Zero і RPi 3*

Розглянемо, як при цьому змінювалася температура процесорів. На рис. 10 відображено залежність завантаження процесора від температури RPi Zero на відстані 5 м. У моменти, коли знижувалося завантаження, то і знижувалася температура — це може означати, що спрацьовував тротлінг — механізм захисту процесора від перегріву (з пропуском тактів).

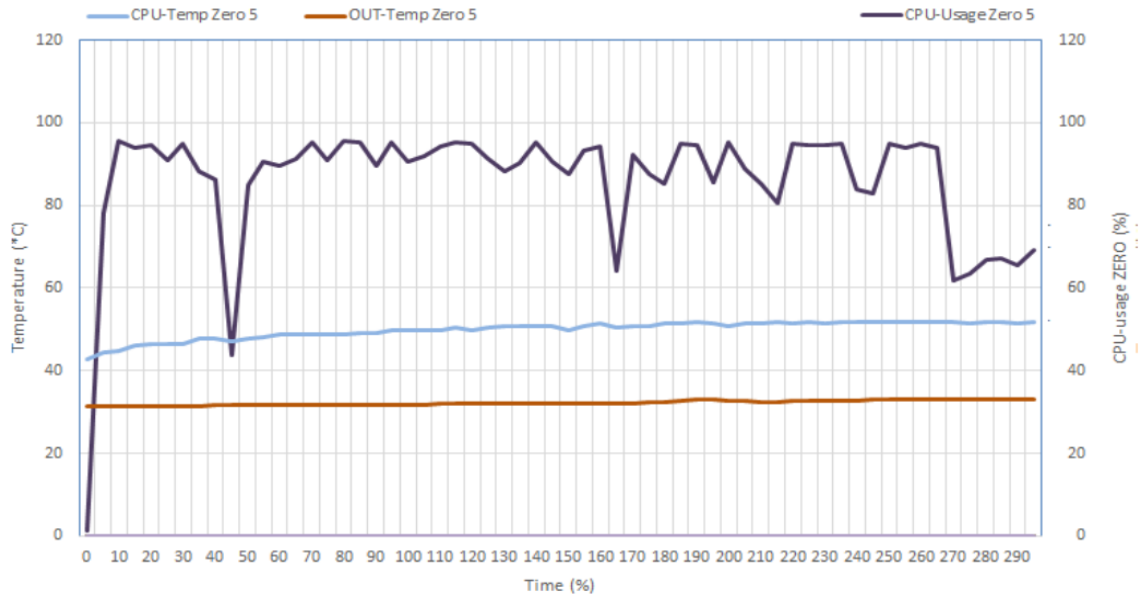

*Рис. 10. Залежність завантаження процесора від температури RPi Zero на відстані 5 м*

На RPi 3 завантаження процесора більш рівне, тому і температура більш рівно трималася (рис. 11, порівн. з рис. 12).

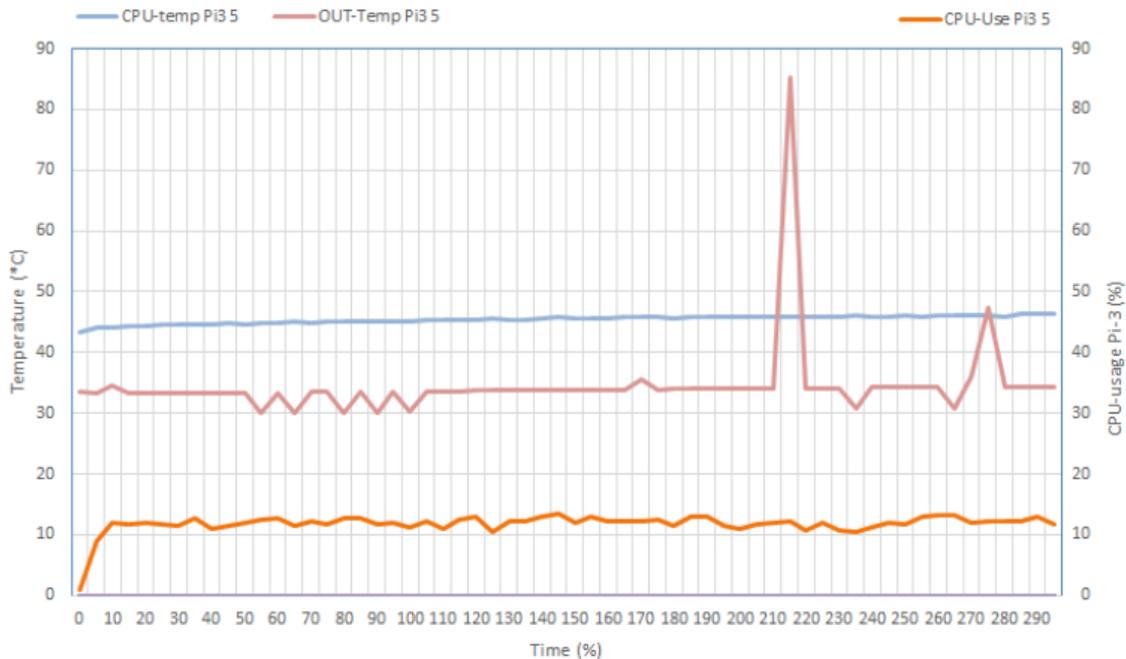

*Рис. 11. Залежність завантаження процесора від температури RPi 3 на відстані 5 м*

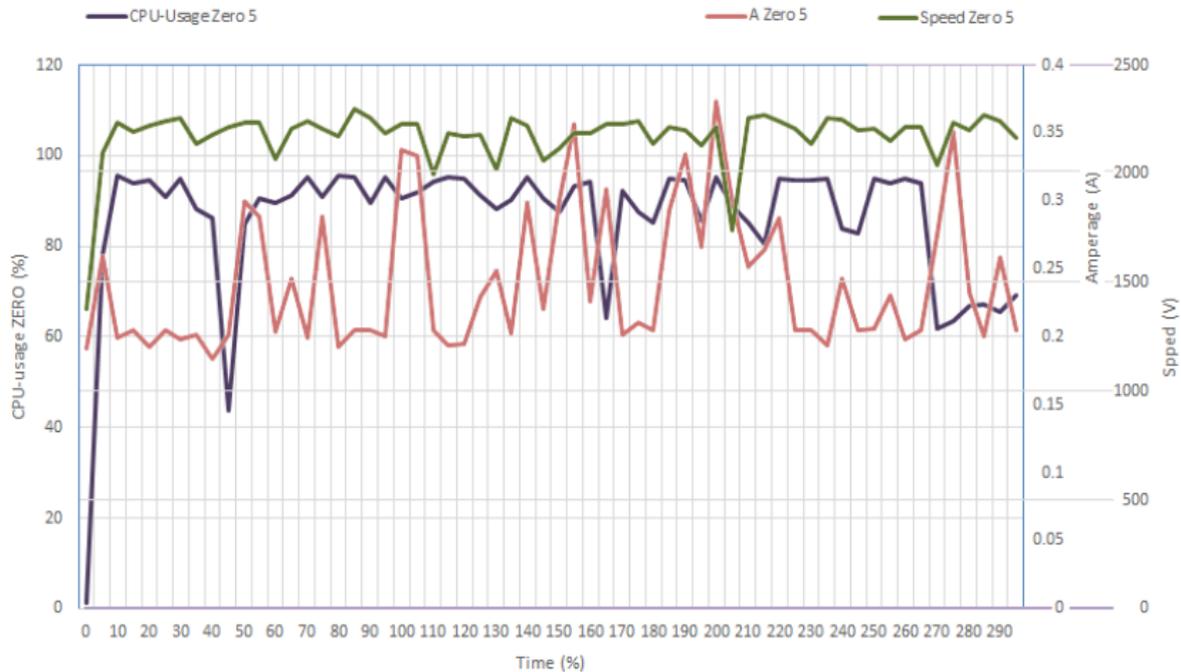

*Рис. 12. Порівняння завантаження процесора, споживання струму і швидкість завантаження файлу з RPi Zero на відстані 5 м*

Найбільш помітно знижувалася швидкість закачування при більш різкому зниженні споживання струму. Споживання струму різко збільшувалася тоді, коли потрібно було знизити завантаження процесора щоб зменшити температуру процесора (рис. 13).

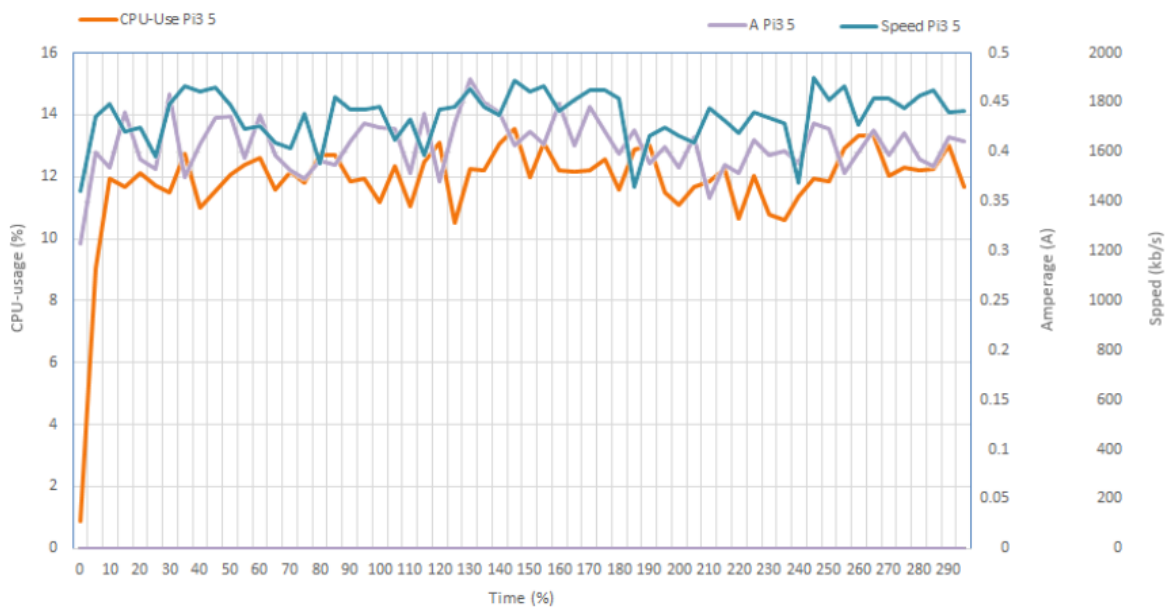

*Рис. 13 – Порівняння завантаження процесора, споживання струму і швидкість завантаження файлу з RPi 3 на відстані 5 м*

За допомогою кореляції Пірсона можна розрахувати залежність факторів між собою:

$$r = \frac{n\sum xy - (\sum x)(\sum y)}{\sqrt{(n\sum x^2 - (\sum x)^2)(n\sum y^2 - (\sum y)^2)}}, \quad (1)$$

де $x$ — значення одного фактору; $y$ — значення іншого чинника до якого відноситься співвідношення.

У табл. 1 наведено результати розрахунку коефіцієнта кореляції Пірсона для різних відстаней за формулою (1). Коефіцієнт кореляції варіюється в діапазоні від плюс 1 до мінус 1. При наявності позитивної кореляції збільшення одного показника сприяє збільшенню другого. При негативній кореляції збільшення одного показника тягне за собою зменшення іншого. Чим більший модуль коефіцієнта кореляції, тим помітніша зміна одного показника відбивається на зміні другого. При коефіцієнті рівному 0 залежність між ними відсутня повністю.

*Таблиця 1*
**Результати розрахунку коефіцієнта кореляції Пірсона**

| RPi Zero / RPi 3 (0,5 м) | | | | | | |
|---|---|---|---|---|---|---|
| | Швидкість | Завант. CPU | Темп. CPU | Зовн. темп. | Напруга | Сила струму |
| **Швидкість** | 1,00/1,00 | | | | | |
| **Завант. CPU** | 0,12/0,05 | 1,00/1,00 | | | | |
| **Темп. CPU** | 0,31/0,24 | 0,53/0,33 | 1,00/1,00 | | | |
| **Зовн. темп.** | 0,47/–0,21 | 0,24/0,16 | 0,84/–0,03 | 1,00/1,00 | | |
| **Напруга** | –0,03/–0,19 | 0,02/–0,36 | –0,11/0,04 | 0,03/0,01 | 1,00/1,00 | |
| **Сила струму** | –0,13/0,01 | 0,15/0,39 | 0,24/0,05 | 0,12/0,27 | –0,03/–0,34 | 1,00/1,00 |
| RPi Zero / RPi 3 (5 м) | | | | | | |
| | Швидкість | Завант. CPU | Темп. CPU | Зовн. темп. | Напруга | Сила струму |
| **Швидкість** | 1,00/1,00 | | | | | |
| **Завант. CPU** | 0,55/0,38 | 1,00/1,00 | | | | |
| **Темп. CPU** | 0,30/0,24 | 0,26/0,46 | 1,00/1,00 | | | |
| **Зовн. темп.** | 0,11/0,02 | –0,02/0,05 | 0,87/0,19 | 1,00/1,00 | | |
| **Напруга** | –0,16/–0,04 | 0,06/–0,17 | –0,18/–0,08 | –0,18/–0,04 | 1,00/1,00 | |
| **Сила струму** | 0,05/0,32 | 0,01/0,43 | 0,22/0,17 | 0,21/–0,07 | –0,12/–0,10 | 1,00/1,00 |

## 7. ВИСНОВКИ ТА ПЕРСПЕКТИВИ ПОДАЛЬШИХ ДОСЛІДЖЕНЬ

Виходячи з отриманих результатів можна зробити висновки, що всі досліджувані фактори, а саме завантаження процесора сервера, температура процесора, споживаний струм впливають на швидкість завантаження файлів. При роботі процесора на пікових значеннях температури виникає процес захисту від перегріву, що називається дроселюванням тактів. У цей час знижується тактова частота процесора і знижується його продуктивність і ефективність. Це тягне за собою зниження швидкості завантаження файлу. За отриманими даними видно, що у RPi Zero W процес дроселювання тактів відбувався з більш високим зниженням частоти тактів і продуктивності процесора ніж у RPi 3.

Для великого обсягу передачі даних по безпроводовій мережі RPi 3 більше підходить ніж RPi Zero W, тому що третя версія більш продуктивна, відповідно вона може більше підтримувати одночасних завдань. Однак для простих завдань більше підходить RPi Zero тому так як ця версія вимагає менше електроенергії. Для IoT пристроїв автономність і споживання електроенергії — це дуже важливий показник.

Проведена робота не вичерпує всієї глибини дослідження. Можливі напрямки подальших наукових досліджень включають більш удосконалений підхід отримання інформації і глибший статистичний аналіз отриманих даних. Даний технічний комплекс у вигляді RPi спільно з датчиком температури DS18B20 і датчиком струму, напруги та потужності INA219 можна інтегрувати в такі системи як системи моніторингу в дата-центрах, системи «розумного будинку» і т. п. Даний технічний комплекс вже використовується для моніторингу температури серверної у дата-центрі. Як приклад, для візуалізації, моніторингу та аналізу отриманих даних використовується платформа Grafana, зображена на рис. 14.

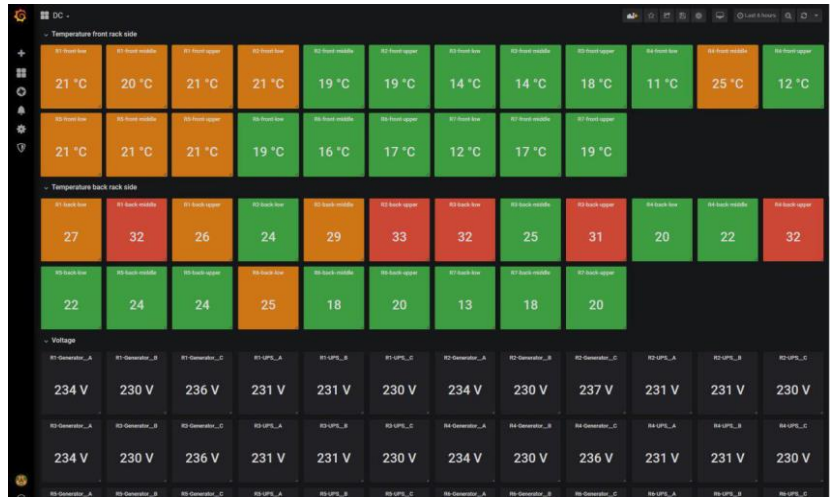

*Рис. 14. Дашборд Grafana із зображенням показників температури та напруги*

## СПИСОК ВИКОРИСТАНИХ ДЖЕРЕЛ

[1]. Raspberry Pi Foundation. (2015, Apr.). "Raspberry Pi 3 Model B." [Режим доступу]. https://www.raspberrypi.org/ products/raspberry-pi-3-model-b/.

[2]. Raspberry Pi Foundation. (2017, Feb.). "Raspberry Pi Zero W." [Режим доступу]. https://www.raspberrypi.org/ products/raspberry-pi-zero-w/ .

[3]. В. Ю. Соколов і Д. М. Курбанмурадов, «Методика протидії соціальному інжинірингу на об'єктах інформаційної діяльності», *Кібербезпека: освіта, наука, техніка*, №1, 2018, сс. 6–16. doi: 10.28925/ 2663-4023.2018.1.616.

[4]. В. Ю. Соколов, «Порівняння можливих підходів щодо розробки низькобюджетних аналізаторів спектру для сенсорних мереж діапазону 2,4–2,5 ГГц», *Кібербезпека: освіта, наука, техніка*, №2, 2018, сс. 31–46. doi: 10.28925/2663-4023.2018.2.3146.

[5]. Oestoidea. (2017, Sept.). "Access Point on Raspberry Pi 3 with Parameter Display." [Режим доступу]. Available: https://github.com/Oestoidea/Adafruit_Python_SSD1306

[6]. Python Software Foundation. (2018, Jun.). "pi-ina219 1.2.0. Project description." [Режим доступу]. https://pypi.org/project/pi-ina219/.

[7]. Les Pounder. (2017, Jun.). "DS18B20 Temperature Sensor With Python (Raspberry Pi)." [Режим доступу]. https://bigl.es/ds18b20-temperature-sensor-with-python-raspberry-pi/

**Volodymyr Sokolov**
MSc, senior lecturer
Borys Grinchenko Kyiv University, Kyiv, Ukraine
ORCID: 0000-0002-9349-7946
*v.sokolov@kubg.edu.ua*

**Bohdan Vovkotrub**
MSc
Borys Grinchenko Kyiv University, Kyiv, Ukraine
OrcID: 0000-0002-6196-4554
*bvovkotrub@gmail.com*

**Yevhen Zotkin**
MSc
Borys Grinchenko Kyiv University, Kyiv, Ukraine
OrcID: 0000-0002-6029-1044
*evgeniy.zotkin199702@gmail.com*

# COMPARATIVE BANDWIDTH ANALYSIS OF LOW-POWER WIRELESS IOT-SWITCHES

**Abstract.** The article presents the research and comparative analysis of the bandwidth of low-power wireless IoT devices as wireless switches. The following IoT devices were investigated: Raspberry Pi 3 Model B and Raspberry Pi Zero W. The DS18B20 and INA219 sensors investigated and analyzed the dependence of FTP multimedia data transmission speed on wireless Wi-Fi network on the temperature of the switch processor, temperature. The environment and the current and voltage consumed by the switch. Advantages of sensors with GPIO interface over analog meters for this experiment are revealed. Much of the work is devoted to the development of automation of results from GPIO interfaces, which helped eliminate human error and get more accurate metrics. Measurement automation was developed using Python 3.7 programming language. Using the INA219 library we were able to obtain current and voltage indicators from the *ina219* board. To get temperature indicators sufficiently built into Python libraries to read temperature files in Raspbian. The article focuses on the synchronicity of measurement results records for more accurate analysis. Therefore, an FTP client was developed that measures the download speed of the file from the FTP server and records the results simultaneously with temperature, current and voltage measurements. To this end, attention is drawn to the multithreading in Python programming language and the transmission of commands using TCP sockets in that language. As a result, the dependence of the measured factors was calculated using the Pearson correlation formula. These measurement factors affect the autonomy and energy consumption, which is very important for IoT devices, and therefore, among the devices tested, recommendations were made regarding their choice when used depending on the conditions.

**Keywords:** bandwidth; wireless network; IoT; INA219; DS18B20; Raspberry Pi; RPi.

## REFERENCES

[1]. Raspberry Pi Foundation. (2015, Apr.). "Raspberry Pi 3 Model B." [Online]. https://www.raspberrypi.org/products/raspberry-pi-3-model-b/ [Aug. 25, 2019].

[2]. Raspberry Pi Foundation. (2017, Feb.). "Raspberry Pi Zero W." [Online]. https://www.raspberrypi.org/products/raspberry-pi-zero-w/ [Aug. 25, 2019].

[3]. V. Yu. Sokolov and D. M. Kurbanmuradov, "Methods of Counteracting Social Engineering at Objects of Information Activity [Metodyka protydiyi sotsial'nomu inzhynirynhu na ob"yektakh informatsiynoyi diyal'nosti]," *Cybersecurity: Education, Science, Technology*, no. 1, 2018, pp. 6–16. doi: 10.28925/2663-4023.2018.1.616.

[4]. V. Yu. Sokolov, "Comparison of Possible Approaches for the Development of Low-Cost Spectrum Analyzers for 2.4-2.5 GHz Sensor Networks [Porivnyannya mozhlyvykh pidkhodiv shchodo rozrobky nyz'kobyudzhetnykh analizatoriv spektru dlya sensornykh merezh diapazonu 2,4–2,5 HHts]," *Cybersecurity: Education, Science, Technology*, no. 2, 2018, cc. 31–46. doi: 10.28925/2663-4023.2018.2.3146.

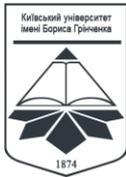

[5]. Oestoidea. (2017, Sept.). "Access Point on Raspberry Pi 3 with Parameter Display." [Online]. Available: https://github.com/Oestoidea/Adafruit_Python_SSD1306 [Aug. 25, 2019].
[6]. Python Software Foundation. (2018, Jun.). "pi-ina219 1.2.0. Project description." [Online]. https://pypi.org/project/pi-ina219/ [Aug. 25, 2019].
[7]. Les Pounder. (2017, Jun.). "DS18B20 Temperature Sensor With Python (Raspberry Pi)." [Online]. https://bigl.es/ds18b20-temperature-sensor-with-python-raspberry-pi/ [Aug. 25, 2019].